# Impact of the Evolution of Smart Phones in Education Technology and its Application in Technical and Professional Studies: Indian Perspective

Manoj Kumar

Assistant Professor, Department of Computer Science,
Maharaja Surajmal Institute, Janakpuri, New Delhi, India
Email: manoj.rke77@gmail.com

## Abstract

*The greatness of any nation depends largely on the system of education that is used to nurture its talent from within. With the digital era taking the spotlight, and the world rapidly reforming into a global village, it is now quintessential that a spirit of healthy competitiveness be inculcated in the budding minds of this country. While trying to remodel and upgrade the education system, a key issue is that of quality of education processes in the country. Needs and expectations of the society are changing very fast and the quality of higher education requires to be sustained at the desired level.*

*The use of internet for educational purposes has increased many folds among Indian youths. Online video lectures and e-books are the emerging trends among learners. The birth of high speed internet access and its availability on recently evolved smart phones has opened several new avenues for learning. The growing popularity of these smart phones among the youth can potentially revolutionize the way we learn. The introduction of 3G technology is already being pinned as the next big thing in the mobile internet revolution.*

*This paper discusses the use of Smart Phones in Education Technology and its application in Technical & Professional studies in India. We intend to put forward some challenges and advices.*

## 1. Introduction

The Indian heritage and culture has always been one of leaning and enlightenment. As is popularly said "Sa Vidya Ya Vimuktaye" (which liberates us is education). The idea of education has been very grand, noble and high since ancient times. Early evidences such as the Vedas and the Upanishads, suggest that education had an esoteric existence. The word Upanishad itself translates into learning acquired by sitting at the feet of the master or the guru. In those times, education was imparted orally to the scholars. Later the Gurukul system of education came into existence. This is the oldest and rightfully the most effective education system mankind has ever known. Gurukuls were the temples of learning that served as the repositories of its philosophical, literary, artistic, scientific & spiritual achievements. They were also a medium of transmission of our heritage to the future generations. In time, several educational institutions took shape all across India. These centers were specialized in imparting higher education. The Nalanda University was the first university-system of education in the world. The universities of Taxila, Ujjain, Benares, Vallabi, Ajanta, Madura and Vikramsila were also very famous in ancient India. These were the fountains of knowledge where students from all across Asia came to quench their thirst for knowledge, truth and enlightenment. With time, India lost its native educational system, rather slowly but surely. This can be attributed to the long history of invasions and attacks by warriors and kings all across the world. The culture shock diluted the importance and relevance of the traditional Indian educational system and ingrained the western educational system into the Indian education fabric. The colonial rule was a key contributor to this cause. The present system of education was founded and introduced by the British in the $20^{th}$ century. By the time India got its freedom from colonial rule in 1947, its education system had undergone a metamorphosis.

During the last 60 years post independence, we have witnessed a number of changes in our education system. Both the regular and distance modes are popular in India today. The exponential growth of the internet is ready to transform the delivery of our education system. India comes only next to China in terms of the number of mobile subscribers globally. Smart phones sales are also at its boom in India. In this paper we have studied the changing scenario in Higher Education and impact of the arrival of smart phones on Indian education system.

## 2. Higher Education Scenario in India

India is a vast country whose engineering and professional student population outnumbers every other country, except possibly China. India has one of the largest higher education systems in the world [9]**.**

Since the early 1950's higher education has been diversified and has extended its reach and coverage quite significantly. At the time of independence, 1947, the size of the higher education system in terms of number of educational institutions, and teachers was meager but since that time there has been an exponential increase in three indicators of higher education, namely the number of educational institutions, teachers and students [10].

Today we have 504 Universities, with varying statutory bases and mandates. Of these, 40 are Central Universities, 243 are State Universities, 130 are Deemed Universities, 53 are State Private Universities, and 33 are Institutions of National Importance established by Central legislation including 5 institutions established under State legislation. We boast of a total teaching faculty of around 0.6 million in higher education [12]. The number of colleges has increased from 500 in 1947 to around 26000, where most of the enrolment in higher education occurs.

In the spheres of technical education, there were a whooping 1346 engineering colleges in India by the end of the year 2004, all of which were approved by the All India Counsel for Technical Education (AICTE) with a seating capacity of 440,000. In 2008, a total of 2388 engineering colleges were registered, with a total student intake capacity of 8.2 lakh students and 1231 management institutes, with an annual intake capacity of 1.5 lakh students. Other than engineering and MBA colleges, there were 1137 MCA, 1001 Pharmacy, 86 HMCT, 11 Applied Arts and Crafts, and 285 PGDM institutes were approved by AICTE. By the end of December 2008, the number of applications pending to seek AICTE approval was 886 for engineering and technology, 250 for MBA, 834 for MCA, 120 for Pharmacy, 124 for HMCT, and 1 for Applied Arts and Crafts - a total of 2237 [13]. The total number of engineering seats has crossed the mark of 1 million in 2009.

The Government of India has announced an increase in the number of world class institutions like Indian Institute of Technology (IIT), Indian Institute of Management (IIM), National Institute of Technology (NIT) and Central Universities (CU) in recent past. These institutes have already marked their names among the top higher educational institutes of the world. SAARC University has also started its functioning from a temporary campus in New Delhi in 2010. The students who will be granted admission in SAARC University will be chosen from the eight nations that are members of the South Asian Association for Regional Cooperation (SAARC)[14]. The Government of India, through the Ministry of Human Resource Development (MHRD), has founded 5 Indian Institutes of Science Education and Research (IISER), specifically to bridge the gap between teaching-only universities and research-only institutes. There is also a proposal to establish 14 World Class Universities (WCU) having various schools including medical and engineering courses [15].

Apart from these world class institutes with excellent facilities, different corporate houses are increasingly lining up to announce their dream projects in the field of education.

A majority of technical and professional institutes which have grown up in the last few years are either self-financing or privately managed. This can safely be attributed to the impact of globalization and some changes in the government policies. Due to the exponential growth of the number of institutions in both the public and private sectors in recent past, India is facing an acute shortage of quality teachers. The top 22 universities in India have 34% vacancy in teaching jobs [16].The overall scenario in other universities and colleges is no different. A majority of teachers are young and inexperienced. India needs many more teachers for effective implementation of higher education in technical and professional courses.

## 3. Higher Education in Distance mode through IGNOU and other Open Universities

Open learning and distance education refer to learning mechanisms, that focus on freeing pupils from constraints of time and place while offering flexible opportunities for erudition. For many students, open and distance learning (ODL) is a way of combining work and family responsibilities with educational opportunities [18].

With an aim to cater the needs of disadvantaged segments of society and to encourage, coordinate and standardize distance and open education in India, "Indira Gandhi National Open University (IGNOU)", was established in 1985, with a budget of INR 2000 crore [19]. The establishment of IGNOU marked the beginning of a new and prestigious era in the history of distance higher education in India. IGNOU has been teaching its students through ODL substantially by the use of Gyan Darshan, an educational TV channel, Gyan Vani, FM Radio channel, interactive teleconferencing via satellite and through Doordarshan DTH. The success of IGNOU can be measured by the fact that it currently serves approximately 3.8 million students hailing from various states of India and about 40 countries abroad [19]. The huge success of IGNOU has led several states in India to announce open universities. Additionally, a few conventional universities have launched several programmes that can be taken through the distance mode.

Typically, ODL is delivered using a variety of print and electronic systems either through synchronous communication (learning in which all parties participate at the same time) or through asynchronous communication (learning in which parties participate at different times). The main systems are mediated by correspondence, audiovisual means (television and radio), multimedia (audio and text files), and the Internet [18]. These channels have co existed, without replacing each other and are used in different combinations in distance education systems across the world.

The target market segment for distance education is varied but is found to mostly comprise of the working adult population, in both professional and in-service education [1].Two key factors have led to an explosion of interest in distance learning: the growing need for continual skills upgrading and retraining; and the technological advances that have made it possible to teach more and more subjects at a distance [20].

In this mode of education the learners have less time for face to face interaction with teachers in comparison of regular mode education. The study centers of the Universities providing distance education can be seen in every nook & cranny of all the major cities across the country. Sadly, these centers fail to provide quality teachers for distance learners.

## 4. Role of Web in Education

The sharp rise in the number of literate youth, the expansion of higher education as a field, the reduction of telecom call rates, the rise in consumer income and the socio-economic changes have increased the demand for internet sharply in the past decade in India. The World Wide Web is host to an ever expanding information explosion. In the absence of quality teachers and enriched libraries, the students demand for surfing the internet for educational purposes has seen a sharp rise. Several institutions have started to provide online study material through their websites. Use of internet in education has a potential to revolutionize the way we live our lives, and the effects have already started making themselves visible [2]. There were around 81 million internet users in India in 2010- a number that will triple by 2015 to 237 million [21] [22]. According to a research conducted by Internet and Mobile Association of India (IAMAI), the School and College students contributed 44% whereas the young men (21-35 years) contributed 28% of the Internet usage in 2009 in India. The same report reveals that a total of 65% searches were made to collect educational information on the web in 2009 and Internet usage has gone up from 9.3 hours/ week in 2008 to 15.7 hours/week in 2009 [23].

India is experiencing an age of transformation in the field of higher education these days. The availability of qualified and quality instructors is essential to ensure the development of Human Capital that caters to the demands of both Industry and Academia. In today's liberalized environment, Government is in earnest need to make the higher education scenario globally competitive. The Parliamentary Standing Committee on HRD in its 172$^{nd}$ report has recommended that we must exploit our ICT potential for its penetration to the Country's remotest corner to expand the access to higher education [24]. To build high speed nationwide communication among Indian Universities, UGC-Infonet program was started by University Grant Commission (UGC). In order to achieve the goal of quality and excellence in higher education, the Ministry Of Human Resources Development, Government of India launched the "National Program on Technology Enhanced Learning (NPTEL)" in the year 2003. The aim of the NPTEL is to develop curriculum based video lectures and web courses to enhance the quality of engineering education in India. The course videos are available in streaming mode and may also be downloaded for viewing offline [25]. The video files are available via the IIT channel in Youtube. Seven IIT's and Indian Institute of Sciences (IISc) have worked together to develop web and video based material. Around 110 video lectures (Approximately 4500 hours) courses were available at the end of phase I. There will be around 400 video (Approximately 16000 hours) lecture courses at the end of phase II. Unlike phase I which offered only undergraduate courses, phase II will see post-graduate courses being offered in 5 out of the 20 disciplines. When this project completes, this will be largest video repository of technical lecture courses in the world in video format and will be helpful to everyone who is interested in enhancing his/her learning[26].

NPTEL has become one of the most popular educational programmes on the internet, spread across 17 countries and registering over four million hits. With fresh approvals from the HRD ministry, the project coordinators plan to offer the equivalent of a degree or a diploma to students enrolled in the Virtual University. By the time NPTEL launch the Open Virtual University in 2012, there will be around 1,000 courses at both the undergraduate and postgraduate level [27].

Course contents of NPTEL are now being utilized for Training of Trainers (TOT) that may further the cause of development of quality professionals. In addition, the course material is freely accessible by everyone independent of their geographic location. Open and distance education using NPTEL is all set to standardize the technical education scenario in the country.

### 4.1. NPTEL not alone: World's top universities offer the same

In addition to the NPTEL, several other offerings exist for the online students. There are a number of web sites that offer thousands of free video lectures and related course materials, including many offerings from major universities. Apple's iTunes U and Google's You Tube EDU have emerged as wonderful platforms for such online video lectures. iTunes U offers an open access to content from world class institutions and universities such as Harvard, MIT, Cambridge, Oxford, Stanford, Yale, Princeton, Columbia, United Kingdom's Open University, University of Melbourne and Université de Montréal. Over 800 universities throughout the world have active iTunes U sites [28]. Interested learners now have access to over 350,000 video and audio files from educational institutions around the world. It has become one of the world's most popular online educational catalog with more than 300 million downloads [29]. On the other end UC Berkeley, University of California Los Angeles (UCLA), University of Phoenix, University of Missouri Kansas City, Stanford, Washington State University, Penn State University, IIT and IGNOU channels have marked their presence among the popular one's on You Tube EDU [30]. iTunes U and You Tube EDU give potentially anyone, the chance to experience university courses, lab demonstrations, sports highlights, campus tours and special lectures while transcending all geographical boundaries. All the contents of iTunes U and You Tube EDU are free and can be enjoyed on a PC, or directly onto an iPhone, iPod and iPad. Besides Apple's iTune and You Tube EDU, a large number of websites provide online video lectures. Teacher Tube [38], LearnersTV [39], Lecturefox [40], Academic Earth [41], Testeachers Online [42], Khan Academy [43], Academic Info [44], FreeVideoLectures [45], and OnlineVideoLectures [46] are few names among the popular ones. Khan Academy is an incredible example of online video lectures managed individually by an MBA scholar from Harvard. He has posted more than 2000 videos on different topics on Mathematics, Sciences, Economics, Business, Finance and History etc., starting from very basics and moving on to college level topics. His website has received more than 20 million hits across the globe. The inclination towards online lectures is increasing, as 40% of the Internet users in India are watching online video lectures and Google now stands as the most popular destination for the learners [50].

### 4.2. Trend towards e-books

Apart from the online video learning, we also observe a recent trend suggesting the popularity of e-learning through e-books. Millions of documents and books are now available to students at the click of a mouse button. An e-book can either be conveniently purchased online or downloaded for free, and hence be used immediately. This is in contrast to the conventional method of purchase whereby a person may either borrow the book or buy the book by visiting a bookshop or chose to visit the local library. All these choices involve time and place constraints that are found to act as major deterrents for the purchase of appropriate books in most students. All the later choices restrict the hours in which a buyer may materialize his purchase and restrict the number of available outlets for purchase. Additionally, when compared to print publishing, e-books are cheaper and easier to share and preserve for a long time. Anyone can read, download and print them instantly 24x7 from any part of the world. New marketing models for e-books have been developed and dedicated reading hardware's are being produced. In the United States, the Amazon Kindle model and Sony's PRS-500 are the dominant e-reading devices. Apple Inc. launched a multi-function device called the iPad. Barnes & Noble Nook is another android e-book reader device that has been developed by the American book retailer Barnes & Noble. The trend of dedicated e-book reader may soon make itself visible in is full glory. Scribd [51], FreeBookSpot [52], Free-eBooks [53], The eBook Directory [54], ManyBooks [55], 4eBooks [56], Globusz [57], FreeComputerBooks [58], FreeTechBooks [59], OnlineComputerBooks [60] are some of the popular web addresses for e-books, in use today. The World Wide Web is flooded with books on diverse subjects. Keeping up with the changing times, Google launched its eBookstore in December 2010, joining many large e-book sellers online. It claims to be the Internet's largest online e-book store. With more than 3 million e-books for sale, it has more e-books than Amazon's market-dominating 2.5 million digital books [61].

All these developments are indicative of the potential of Internet- supported learning to provide quality education at an affordable cost and in a convenient form. Additionally, increase in online programs, courses, or class sessions have witnessed a related decrease in the need for physical facilities, thus enabling service of more students without any additional costs [62].

## 5. Impact of Smart Phone on Distance Learning

In Section 4, we discussed that the availability of online learning is gaining popularity in India. NPTEL is one of the best examples in this direction. Use of Internet has become a part of life of every student. These days, use of mobile phones for internet purposes has become a habit with all students. According to Vinay Goel, head of products, Google India, a total of 40 million users access the internet through mobile phones in India. He further estimates that by 2012, the number of mobile internet users will surpass those entering the net via their laptops or desktops [63]. It has been observed that 55% of the 'mobile only[*]' are the students in India [64]. Fixed-line internet is being completely bypassed by these mobile users and it is also an exciting time for content owners and brands to interact with their customers. India is the second largest telecommunication network in the world in terms of number of wireless connections after China with more than 752 million mobile phone subscribers in December, 2010 [65].Almost all the telecom service providers in India have launched low budget lucrative offerings to access internet over mobile phones. Aircel, Airtel, Tata Docomo, Uninor, Videocon and other telecom operators offer mobile internet startup plans within INR 100. These offerings are very popular among the students. These plans give unlimited internet access for 30 days. Students usually connect their mobile phones to their laptop/desktop to access the internet. Hence we observe that there already exists a consumer behavior that is conducive to the digitization of distance learning initiatives.

---

[*] *The mobile only generation is those who use the mobile internet but have either never used or only use the desktop internet once a month or less.*

In recent times, smart phones have gained remarkable popularity in consumer markets across India. India today serves as a lucrative market for all mobile phone manufacturers across the world. Apart from the big players like Nokia, Apple, RIM, HTC, Samsung, LG, Motorola and Sony Ericsson the Indian mobile handset makers Lava, MicroMax, Spice, Karbonn, Videocon and Intex have flooded the Indian mobile market with wide variety of mobile handsets.

The popularity of smart phones is growing and people have started to use their smart phones to access the internet directly over their mobile phones. According to IDC India the smart phones sale was expected to touch 6 million units by end of calendar 2010 in India [66].

According to Scott Steinberg editor Digital Trends-
"A smart phone is essentially a computer in your pocket. It is a cellular phone that does more than just make calls to the point that it can actually serve as a functional laptop or desktop replacement"

## 5.1. Market Potential of Smart Phones

- Today's information workers demand the flexibility to balance work, home and leisure. The smart phones enable them to lead the $21^{st}$ century lifestyle.

- These days' smart phones are being used primarily by consumers in the youth segment who love to access the web, interact through virtual social networks, receive and compose e-mail instantly, check the news and even use utilities such as the GPS navigation as a lifestyle choice.

- Wi-Fi connectivity helps to connect the internet via the hotspots when users are at the airport or other important public places. HD video player allows watching movies with high clarity.

- The popularity of smart phones has created a wave in development of mobile friendly websites. More than 100 million people actively use Facebook from their mobile devices every month [67] whereas mobile browser opera mini has more than 90 million users [68].

- Just a few years back, smart phones were more of a status symbol, but now they have become a must-have productivity aid, literally carrying a lot of information all in pocket.

Today's smart phones are getting faster day by day while providing a high storage capacity. One flagship device, the Motorola Atrix, won the Best Smart phone award in Consumer Electronics Show (CES) 2011 in Las Vegas. The 4G powered Atrix runs on Google Android 2.2 Froyo operating system. It has a 4-inch QHD display that provides maximum multimedia entertainment on 4G networks. This smart phone is powered with dual-core 1GHz Nvidia Tegra 2 SoC processor and 1GB RAM. Qualcomm, a world leader in 3G and next-generation mobile technologies recently announced its next mobile processor architecture for the Snapdragon family. It is all set to redefine performance for the industry, offering speeds of up to 2.5GHz per core and delivering 150 percent higher overall performance, as well as 65 percent lower power than currently available ARM-based CPU cores. These chipsets will be available in single, dual and quad-core versions and include a new Adreno GPU series with up to four 3D cores, and integrated multi-mode LTE modem. Several other devices are pinned to follow suit.

In a nutshell, smart phones or mobile devices will soon become the dominant computing platform for humanity. Morgan Stanley Research estimates sales of Smart phones to exceed those of PCs in 2012. Gartner expects over 500 million Smart phones to sell in 2012.

## 5.2. 3G is Ready to Accelerate

In 2008, India entered the 3G arena when Government owned Bharat Sanchar Nigam Limited (BSNL) launched its 3G enabled mobile and data services. Later Mahanagar Telephone Nigam Ltd (MTNL) also launched its services in Delhi and Mumbai. The private sector service providers such as Tata Docomo, Reliance Communications, Airtel, and Vodafone have also launched its 3G services.

3G is the next generation of mobile communication system. It enhances services such as multimedia, high speed mobile broadband, thus equipping the average mobile user with the ability to watch live TV on his/her mobile handset. One can also enjoy services such as live streaming, download of videos for educational or leisure purposes, news, current affairs and sport content and video messaging all in addition to the usual voice calling facility. Practically mobile users are all set to get a broadband experience with speeds better than 384 KBPS. The connectivity speed can easily be compared to those of the wire line broadband. The biggest benefit of 3G to a learner is the ability to enjoy high speed internet and data service even while on the move, all with the purchase of an affordable 3G handset and subscription to a suitable 3G plan.. One can also utilize Bluetooth, Infrared or data cables to connect other devices such as laptops and net-books, to the internet, while on the move. There also exists the facility of 3G data cards that can be used with computers to access high speed 3G networks.

According to a recent forecast from the Wireless Intelligence, a service of trade group GSMA Ltd., India is all set to have 150 million 3G connections by the year 2014.

## 6. Conclusion and Future Scope

Globalization and technology related developments are 'change drivers' that have significantly re-shaped the landscape of the higher education. New missions and responsibilities assigned by governments in pursuit of national 'wealth creation' and international competitiveness, shrinking public funds, increased need for flexible, lifelong learning arising from the changed nature of work, new learning paradigms and the entry of technology-based new educational providers have had far-reaching effects on higher education.

The impact of the changes in open and distance education institutions are also profound. The difference between traditional universities and distance education institutions has disappeared. The need for lifelong learning and rapid developments in ICT have led many traditional universities to become involved with online delivery, and the commercial potential has attracted many new technology-oriented private as well as public providers. Mobile learning may be used to access the educational opportunities to different segments of the society where distance or other obstacles present a barrier to accessing formal learning centers and to enhance the quality of learning and continued professional development.

The growing demand of smart phone and high speed mobile browsing is ready to change the basics of higher education delivery system. People feel a bonding towards their mobile phones. The services and functionalities provided by a mobile phone are available at all times in both everyday routines and in our special moments. However, the cost of a smart phone, network coverage in remote areas and awareness of the educational contents on web may be few barriers in Indian perspective. The pace at which the mobile subscribers are growing in India, it is evident that mobile phone usage in education is here to stay. While not a panacea for the educational system in our country, the smart phones could be one way to engage and motivate student learning.

# 7 . References